\documentstyle[pra,aps,epsfig,multicol]{revtex}
\newcommand{\epsfxset}{\epsfxsize=7cm}

\newcommand{\modify}[1]{\typeout{NOTE: #1}}

\newcommand{\I}{{\mathrm{i}}}
\newcommand{\vect}[1]{\mbox{\boldmath{$#1$}}}
\newcommand{\AT}{{\vect{A}(t)}}
\newcommand{\R}{{\vect{r}}}
\newcommand{\PSIRT}{{\Psi(\R, t)}}
\newcommand{\llimSum}[1]{{\sum \limits_{#1}}}
\newcommand{\YLOTF} {{\mathrm{Y}_{lm}(\theta, \phi)}}

\begin{document}

\draft

\title{Time-dependent calculation of ionization in Potassium at
 mid-infrared wavelengths}

\author{P.~Maragakis$^{1,2}$\footnote{e-mail:{\tt Maragakis@mpq.mpg.de}},
  E.~Cormier$^{3,4}$ and P.~Lambropoulos$^{1,2}$}

\address{$^1$Department of Physics, University of Crete, P.~O.~Box 2208,
  Heraklion 71003, Greece, and 
  Foundation for Research and Technology-Hellas,
  Institute of Electronic
  Structure \& Laser, 
  P.~O.~Box 1527, Heraklion 71110, Crete, Greece}

\address{$^2$Max-Planck-Institut f\"{u}r Quantenoptik,
  Hans-Kopfermann-str.\ 1, D-85748 Garching, Germany,
  Tel:+49-89-32905-705, fax:+49-89-32908-200}

\address{$^3$Centre d'Etudes de Saclay,
        Service des Photons, les Atomes et les Mol\'{e}cules, DRECAM, 
        F-91191 Gif-Sur-Yvette, France} 
 
\address{$^4$ Centre Lasers Intenses et Applications
        CELIA, EP 2117, 
        Universit\'{e} de Bordeaux I, 
        351 Cours de la Lib\'{e}ration, 
        F-33405 Talence CEDEX, France}

\date{\today}

\maketitle

\begin{abstract}
  We study the dynamics of the Potassium atom in the mid-infrared,
  high intensity, short laser pulse regime.  We ascertain numerical
  convergence by comparing the results obtained by the direct
  expansion of the time-dependent Schr\"odinger equation onto
  $B$-Splines, to those obtained by the eigenbasis expansion method.
  We present ionization curves in the 12-, 13-, and 14-photon
  ionization range for Potassium.  
  The ionization curve of a scaled
  system, namely Hydrogen starting from the 2s, is compared to the
  12-photon results.  In the 13-photon regime, a dynamic
  resonance is found and analyzed in some detail.  The results for all
  wavelengths and intensities, including Hydrogen, display
  a clear plateau in the peak-heights of the low energy part
  of the Above Threshold Ionization (ATI) spectrum, which scales with
  the ponderomotive energy $U_p$, and extends to $(2.8 \pm 0.5) U_p$.
\end{abstract}



\begin{multicols}{2}
\section{Introduction}

The motivation for this work has its origin in recent experimental
data by DiMauro et al~\cite{sheehy:1998:ema} 
who studied high order harmonic generation
(HOHG) and above threshold ionization (ATI) in potassium driven
by strong radiation in the wavelength range 3200--3900 nm.
Although HOHG and ATI have been and continue to be studied 
extensively, the bulk of the data and theory have concentrated on
the noble gases.  Experimental convenience has been one of the 
reasons for this preference, but, at least as far as HOHG is
concerned, their relatively high ionization potentials and 
resistance to ionization have also tilted attention in their
direction.  The alkali atoms belong to an entirely different
class, when it comes to their behavior under strong field
excitation.  For atomic numbers comparable to the respective
noble gas (potassium versus argon in our context), their 
ionization potential is lower by more than a factor of three.
Their excited states and distribution of the oscillator strength
for transitions from the ground state are also considerably
different.  The energy of the first excited state in potassium
is much closer to the ground state than it is in argon.
As a consequence, if we consider, for example, 12-photon
ionization in potassium, five of the photons reach above the first
excited state, and the rest seven must be absorbed within the manifold
of its excited states.  In contrast, for 12-photon ionization
of argon, it takes nine photons to reach the energy range of
the first excited state and only the remaining three will involve excitation
within the manifold of excited states.  In addition, the 
wavelength needed for potassium (about 3000 nm) 
is longer by a factor of three than that needed
for argon
(about 1100 nm).

One might thus expect that, in the process of ionization, an
extensive manifold of excited and Rydberg states will be 
strongly driven and perhaps populated.  This should lead to
a structure in the ATI energy spectrum different in appearance
from what we are accustomed to.  One might also anticipate
that the behavior should have similarities with that observed
in Rydberg states driven by microwave fields.  Resolution 
requirements in photoelectron energy analysis do not allow the
observation of individual ATI peaks in that case, although it
is quite feasible to resolve such peaks in potassium driven
from its ground state by radiation at mid-infrared wavelengths;
as Sheehy et al.~\cite{sheehy:1998:ema}
 have shown.  It is the exploration of 
possible links and similarities between these two situations that
induced us to undertake this work.  Clearly, the requirements
on intensity for saturation are expected to be lower in 
potassium than in argon.  Moreover, given the expected 
participation of manifolds of excited states in potassium, the Keldysh
parameter as a criterion for the departure from multiphoton
ionization may not be as valid as it should be in argon where
most of the energy interval from the ground state to the 
continuum is empty of excited states; imitating thus better the
Keldysh model which is essentially based on a ground state
and an ionization threshold.

The outline of this paper is as follows: The theory, namely the model
used to describe the atom, and the two propagation methods used to
solve the time-dependent Schr\"odinger equation are briefly presented
in section~(\ref{sec:Method}).  Section~(\ref{sec:Results}), starts
with a presentation of the parameter range of the simulations that
follow and a demonstration of convergence by comparing two different
methods.  It then moves on to present results in the 12-, 13-, and
14-photon ionization range, and discuss a low-energy plateau in the
ATI spectrum.  12-photon ionization of Hydrogen starting from the 2s
is compared to the results from Potassium.  We conclude in
section~(\ref{sec:Conclusions}), by summarizing the main findings of
our results.  In the appendix we present some details of the atomic
structure model, and compare it with an alternative approach.



\section{Method}
\label{sec:Method}

Potassium, being an alkali atom, can be considered a single
electron system for most of the phenomena in which double excitation
does not play an essential role.  The first ionization threshold is
about 4.34 eV above the ground state; 18.8 eV are needed to
reach the lowest doubly excited state, which leaves us with enough
room to study single electron dynamics.  The simplest way to do this
is by using a model potential that incorporates the effect of the
core electrons, and thus reduces the dynamics to a single electron
scheme.  Different implementations of this basic idea have been used
so far, with some success, in the Single Active Electron approach,
pioneered by Kulander~\cite{Kulander:1988:tdt}, the frozen core
calculations~\cite{tang:1991:ntd}, and other model potential
calculations~\cite{lambrecht:1998:pca}.  For the purpose of studying 
the dynamics in the mid-infrared, it is sufficient to use a
simple form of the potential, proposed by
Hellmann~\cite{szasz:1985:pta}, which in atomic units is given as:
\begin{equation}
  V_m = - \frac{1}{r} + A \frac{e^{-K r}}{r},
  \label{eq:hellmann:definition}
\end{equation}
where $A$ is 1.989 and $K$ is 0.898.  All formulas that follow are given
in atomic units.  The Hellmann potential with the above mentioned
parameters, results in a ground state energy of 38950 wavenumbers from
the first ionization threshold, which is lower than the energy of
the actual
ground state (35010 wavenumbers) by 11\%.  Owing to the difference between the
ground state energies of the model potential and of the real atom, we
also use a scaled wavelength, namely we rescale the energy of the
photon needed in an experiment that studies the same
process, by the ratio of the model to the theoretical ground state
energy.  An alternative approach is to correct the energy of the
ground state, and probably some matrix elements to satisfy the
oscillator strength sum rules, but this leads to nonlocal
modifications in the Hamiltonian and is difficult to implement
effectively when the propagation is not done in an eigenbasis expansion.

The dynamical part of the problem is treated by solving the resulting
time-dependent Schr\"odinger equation in the dipole approximation:
\begin{equation}
  \label{schroedinger_equation}
  \I \partial_t \Psi(\vect{r}, t)
  =
  \left[
    H_a + D(t)
  \right]
  \Psi(\vect{r}, t),
\end{equation}
where $\Psi$ is the wavefunction describing the outer electron, and
depends on the spatial electronic coordinates and on time, $H_a$ is
the time-independent field-free atomic Hamiltonian, and $D(t)$ is the
dipole interaction of the atom with the field.  We only use the
velocity form of the interaction operator, following detailed studies
on the convergence properties of the solution~\cite{cormier:1996:ogg},
which have shown that the expansion of the wavefunction in terms of
spherical harmonics can be shortened dramatically if the velocity
gauge is used instead of the length gauge.  In the velocity gauge, the
dipole operator can be cast in the following form:
\begin{equation}
\label{DT}
  D(t) = - \alpha \AT \cdot \vect{p},
\end{equation}
where $\alpha$ is the fine structure constant, $\vect{p}$ is the
momentum operator, and $\AT$ is the vector potential which, within the
dipole approximation, has no spatial dependence.  We choose a
convenient form for the pulse envelope, namely a $\cos^2$, avoiding
the long tails of a Gaussian that make the numerics more difficult,
without significantly affecting the results.  The explicit form is:
\begin{equation}
  A(t)=
  \frac{{\cal E}_{0}}{\omega}
  \cos(\frac{\pi t}{\tau})^{2}
  \cos(\omega t),
  \quad
  \text{with }
  t \in [\frac{-\tau}{2},\frac{\tau}{2}],
\end{equation}
where $A(t)$ is the amplitude of the vector potential, $\tau/2$ is the
Full Width at Half Maximum (FWHM), ${\cal E}_{0}$ and $\omega$ the
maximum field strength and fundamental frequency respectively.  
The solution of the resulting time-dependent equation is written in a
system of 
spherical coordinates and expanded in terms of radial functions
and angular spherical harmonics. This choice is dictated by
the central symmetry of the atomic system and has the advantage of
requiring the discretization of only one coordinate. Note that this
choice leads to efficient algorithms only if the linearly polarized
field is not too strong (compared to the coulomb field) in which case
the global symmetry of the entire system would rather be
cylindrical. Writing the solution in the cylindrical system would
require discretization of 2 coordinates~\cite{maragakis:1997:tdh} greatly increasing the
numerical cost of the algorithm. The problem is therefore treated
within "a box" (a sphere in the present case) whose radius is chosen
sufficiently large to contain the expanded atom during the
interaction.  Part of the 
procedure  for testing convergence consists in ascertaining 
that the ATI spectrum is
insensitive to the radius of the box.

We have used two methods of propagating the TDSE, namely a propagation
onto eigenstates in a
box~\cite{lambropoulos:1998:tea,lambropoulos:1998:asi}, and a
propagation on a $B$-Splines
basis~\cite{cormier:1996:ogg,cormier:1997:ati}. 
The expansion of the time-dependent wavefunction on an eigenbasis set
reads:
\begin{equation}
\label{eq:psi_basis}
\Psi (t) =\sum_{l,n} b_{l,n} (t) \Phi^{l}_{n(E)},
\end{equation}
where $\Phi^{l}_{n(E)}$ are the field-free box eigenstates of the atom of
angular momentum $l$.  Since we use linearly polarized light, we only need
the $m=0$ magnetic quantum number, as the initial state has
$m=0$, and dipole transition selection rules forbid mixing of other
magnetic sublevels.  The time-dependent Schr\"{o}dinger equation is
transformed into:
\begin{equation}
\mbox{i} \frac{d}{dt}b_{l,n}(t)=\sum_{n^{\prime}l^{\prime}}(E_{nl}
\delta_{nn^{\prime}}\delta_{ll^{\prime}}-D_{nl,n^{\prime}l^{\prime}}(t))
b_{l^{\prime},n^{\prime}}(t),
\end{equation}
with the initial condition $|b_{l=0,n=1}(0)|^{2}=1$.  $E_{nl}$ are the
eigenvalues in the box; $D_{nl,n^{\prime}l^{\prime}}$ are the dipole
matrix elements.  Thus the problem has been transformed to a set of
ordinary differential equations for the coefficients $b_{l,n}(t)$ of
the wavefunction, which are solved using a high order, explicit
propagation technique, namely a fifth-order and sixth-order
Runge-Kutta-Verner method.
The ionization yield is calculated by adding up
the occupation probabilities of all discretized continuum states at the end
of the pulse; the above threshold ionization (ATI) spectrum is
obtained by the window operator projection
technique~\cite{kulander:1992:tds,gavrila:1992:ail}.  Bound state
populations are given directly by the square of the norm of the
coefficients of equation~(\ref{eq:psi_basis}).  For the construction
of the box-eigenstates that are used as our basis, we use an
expansion onto $B$-Splines, a method that is gaining momentum in many
parts of atomic physics as was pointed out
in~\cite{sapirstein:1996:ubs}.  The codes we use are based on ideas
published in~\cite{tang:1991:ntd,lambropoulos:1998:tea}, 
which have been expanded to accommodate the need for large boxes.
It should be stressed that after the basis has been
constructed (i.e.\ energies and matrix elements have been
calculated), the rest of the procedure is neutral to the 
technique for the construction of the basis.

The second approach rests on expanding the radial part of the
time-dependent wavefunction directly onto
$B$-splines~\cite{cormier:1996:ogg,cormier:1997:ati}: 
\begin{equation}
  \label{eq:Psi_BSplines}
  \PSIRT
  =
  \llimSum{i l} b_i^{l}(t) \frac{B_i^{(k)}(r)}{r} \YLOTF
\end{equation}
where in addition to the spherical harmonics, $B_i^{(k)}(r)$ is the
$i$-th $B$-spline of order $k$ depending only on the radial coordinate
and $b_i^{l}(t)$ are time-dependent coefficients to be determined by
the solution of the TDSE.  Again, only $m=0$ magnetic quantum numbers
are relevant.  The major difference in this approach is that
we need not prediagonalize anything other than the initial state.
Substitution of equation~(\ref{eq:Psi_BSplines}) into the
Schr\"odinger equation~(\ref{schroedinger_equation}) leads to a banded
system of differential equations, which is solved by implicit
propagation techniques, currently involving a Bi-conjugate gradient
method with preconditioner.  This second method of propagation scales
only linearly with increasing box size: it is thus more efficient
when we need a large box.  Although our original exploratory
calculations have been made with the eigenbasis expansion method, which
works quite well for small boxes, most of the results presented in
what follows have been obtained through the direct expansion of the
time-dependent
Hamiltonian onto $B$-Splines.



\section{Results \& Discussion}
\label{sec:Results}

\subsection{General Considerations}
\label{sec:sub:general_considerations}


The only guideline as to what to expect in our 
study are the experimental data by Sheehy et
al.~\cite{sheehy:1998:ema}, in which 12-, 13-, and 14-photon
ionization of Potassium has been studied, with 3.2 $\mu$m, 3.6 $\mu$m,
and 3.9 $\mu$m, 1.5 ps pulses respectively, and intensities close to
saturation.  The 1.5 ps pulse is computationally impractical 
in the TDSE
framework, thus we have chosen to place our study in the short pulse
regime, with the total pulse duration $\tau$ of the $\cos^2$ pulse
being 20 optical cycles, which (depending on the wavelength)
corresponds to a width of 96 fs to 136 fs for the results that we 
present.  The wavelengths we use are scaled, to compensate for
the inaccuracy of the ground state energy, and are such that 12-, 13-
and 14-photon ionization takes place.  An intensity range estimate is
obtained by calculating the generalized cross-section through a scaling
technique~\cite{lambropoulos:1985:mmi,lambropoulos:1987:u}, using the
energy (0.295 Hartree), and radius (5.24 a.u.) obtained with the 
general Hartree-Fock code published by
Froese-Fischer~\cite{fischer:1987:ghf}.  From the cross-section, we
obtain a 
saturation intensity estimate by solving $\Gamma
t_{\text{eff}} = 1$, where $\Gamma$ is the ionization width, and
$t_{\text{eff}}$ an effective pulse duration, of the order of its
FWHM~\cite{charalambidis:1997:mis}.  After the first time-dependent
calculations, it was established that scaling was underestimating 
saturation intensity
by more than an order of magnitude.  We also
calculate the Ammosov, Delone, Krainov (ADK) rate of tunneling
ionization~\cite{ammosov:1986:tic}, which does not depend on the
wavelength.  Solving $\Gamma t_{\text{eff}} = 1$, we
obtain a saturation intensity estimate of about $4 \times 10^{12}$
W/cm$^2$, in agreement with the results of the simulations.

For some characteristic wavelengths, corresponding to 12-photon
ionization scaled from the experimental wavelength, and the limits of
the 13-photon ionization range, we show, in
table~(\ref{tab:parameter:all}), the FWHM duration $\tau/2$, the
scaled-theory saturation intensity I$_{\mathrm{s}}$, the ADK theory
saturation intensity estimate I$_{\mathrm{ADK}}$, and the upper
limit E$_{\mathrm{c}}$ of the converged ATI spectrum 
for a 3000 atomic units box.
E$_{\mathrm{c}}$ is calculated by estimating the energy needed for a
free electron originally placed at the nucleus to reach the boundary
of the box during the pulse; it is a useful simple estimate of the box
size needed to study ATI spectra, as has been shown in section (5)
of~\cite{cormier:1997:ati}.  Next, and for two intensities, we show
the ponderomotive energy $U_p$, which is the major component of the
shift of the Rydberg states and the
continuum~\cite{avan:1976:ehf,mittleman:1984:kmi,freeman:1986:pea}.
It is given by:
\begin{equation}
  U_p = \frac{I}{4 \omega_0^2} \sim I \lambda^2
  \label{eq:ponderomotive}
\end{equation}
where $I$ is the laser intensity, $\omega_0$ the photon energy and
$\lambda$ the corresponding wavelength.  For the highest intensities
that we use, 
the ponderomotive energy is a multiple of the photon energy, which is
around a third of an eV for the wavelengths in the table.  We also
present the Keldysh tunneling parameter
$\gamma$~\cite{keldysh:1965:u}, defined by:
\begin{equation}
  \label{eq:gammasqr}
  \gamma^2 = \frac{I_p}{2 U_p}
\end{equation}
where $I_p$ is the ionization potential, and $U_p$ the ponderomotive
potential.  Note that for all wavelengths, and for intensities up to
$2 \times 10^{12}$ W/cm$^2$, $\gamma$ is larger than one.  Thus,
according to the Keldysh theory of tunneling ionization, the process
lies in the multiphoton regime, and it is meaningful to refer to
the order of the transitions involved.
\end{multicols}

\begin{table}
\caption{Parameters for some of the calculations in Potassium.}
\label{tab:parameter:all}
\begin{tabular}{ddccdcdd}
$\lambda$(nm) & $\tau/2$(fs)& $I_{\mathrm{s}}$(W/cm$^2$) &
$I_{\mathrm{ADK}}$(W/cm$^2$) &
$E_{\mathrm{c}}$(eV) & $I$(W/cm$^2$) & $U_p$(eV) & $\gamma$ \\
\hline 
2880 & 96 & $1.5 \times 10^{11}$ & $4.1 \times 10^{12}$ &7.76 & $10^{11}$ & 0.073 & 5.23 \\
&    &               &      & & $10^{12}$ & 0.733 & 1.65 \\
3125 & 104 & $1.3 \times 10^{11}$ & $4.0 \times 10^{12}$ & 6.59 & $10^{11}$ & 0.086 & 4.82 \\
&     &               &     & & $10^{12}$ & 0.863 & 1.52 \\
3300 & 110 & $1.2 \times 10^{11}$ & $4.0 \times 10^{12}$ &5.91 & $10^{11}$ & 0.096 & 4.56 \\
&     &               &     & & $10^{12}$ & 0.961 & 1.44 \\
\end{tabular}
\end{table}

\begin{multicols}{2}

\begin{figure}
  {\epsfxset \epsfbox{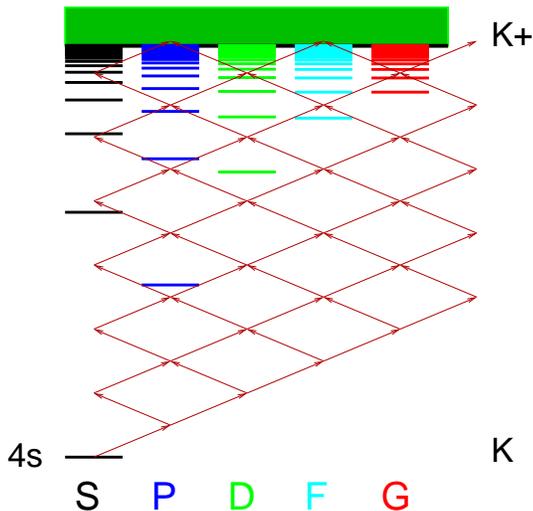}}
  \caption{Truncated atomic structure of Potassium (calculated by the
    Hellmann potential), and quantum paths leading to 13-photon ionization}
  \label{fig:level-photon:3300nm}
\end{figure}
In figure~(\ref{fig:level-photon:3300nm}) we show a visual
representation of Potassium in a 3300 nm field, corresponding to the
lowest energy photons in the 13-photon ionization range.  A truncated
part of the bound atomic levels, of angular momentum up to 4, is
shown, together with the quantum paths leading to 13-photon
ionization.  Graphs of this type prove to be useful tools in the 
qualitative analysis of the processes involved in multiphoton
phenomena.

\begin{figure}
  {\epsfxset \epsfbox{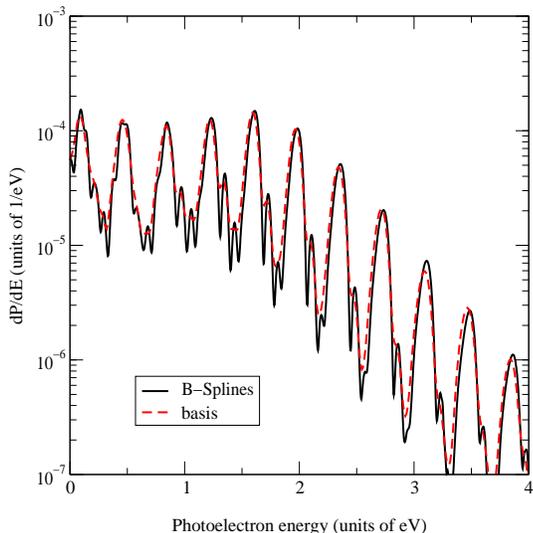}}
  \caption{Two different, time-dependent methods of calculation of the
    dynamics of the Potassium model potential, for a 3300 nm, $7
    \times 10^{11}$ W/cm$^2$, 110 fs, $\cos^2$ pulse.  The full line
    is the ATI spectrum obtained through the $B$-splines expansion
    method, and the dashed line is the ATI spectrum obtained through
    the eigenbasis calculation.}
  \label{fig:ATI:basis-bsp:compare:7e11}
\end{figure}

In our study, we analyze the wavefunction at the end of the pulse, and
thus obtain information on the electron spectrum and ionization.
Depending on the physical quantity we are looking for, the parameters
needed to achieve convergence vary, making it easier to obtain the
value of angle and energy integrated ionization, than the ATI
spectrum.  We have ensured the convergence of our results, by varying
the box size and the grid sampling density, in a way similar to what
has been presented in~\cite{cormier:1997:ati}, and by relying on
empirical findings such as the definition of E$_{\text{c}}$, or the
needed density of discretized continuum states per photon energy, to
guide our parameter choice.  We have in addition conducted a
further independent 
test of the numerics involved, by comparing the
two different methods, i.e.\  the expansion in terms of
box-eigenstates, or the 
direct expansion of the radial part in a
$B$-Splines basis.  A sample result, for a demanding quantity such as
the ATI spectrum in the 13-photon range, is shown in
figure~(\ref{fig:ATI:basis-bsp:compare:7e11}), where the results of
the two simulations at 3300 nm, $7 \times 10^{11}$ W/cm$^2$ and 110 fs
are displayed on top of each other.  The direct $B$-Splines
method involves a box of 3000 a.u., with 3000 linearly sampled $B$-splines
of order 7, for each angular momentum up to $l=20$, parameters that
have proven more than sufficient for the method to converge in the 
range shown.  After the propagation is over, a projection to
scattering states is used to obtain the ATI spectrum.  The eigenstates
expansion involves a box of 2500 a.u., with 2500 linearly sampled
$B$-Splines of order 9 for each angular momentum up to $l=15$ for
constructing the eigenbasis.  This basis was then truncated to the lowest
1000 basis elements per angular momentum; using only the 1000 lowest
states proved to be sufficient for the energy range presented here,
since the higher energy discretized continuum states play a role
in the high energy, low-yield part of the spectrum.  The
window-operator technique~\cite{kulander:1992:tds} was used to obtain
the photoelectron spectrum after the simulation, and the spectrum was
renormalized for the comparison with the $B$-Splines spectrum.

Despite the differences in the methods used, we observe an excellent
agreement of the results in
figure~(\ref{fig:ATI:basis-bsp:compare:7e11}), even on a logarithmic
scale.  The $B$-Splines method shows richer structure at the minima
between the peaks, demonstrating its superiority at the finest parts
of the results.  In detailed analysis of the ATI results (too long to
be
presented here), we have studied the behavior of the side peaks with
intensity and we have noted that different peaks show different shifts with
intensity, which helps us in classifying them as either Freeman
resonances~\cite{freeman:1987:ati} or Bardsley 
fringes~\cite{bardsley:1988:u,cormier:1996:ogg}.
Note further the clean formation of a plateau in the
ATI-peak heights, showing up in both methods, and extending over the
first 5 to 6 peaks; we study this plateau later in the paper.
Since, in theory, the two methods are related by a unitary
transformation of the basis from a spatially localized ($B$-Splines)
to a field-free diagonal (eigenbasis) representation, the agreement of
the results is expected; however, practice has shown that convergence
is strongly 
affected by the underlying numerics, especially when it comes to ATI
spectra that stress the subtle parts of our codes.  
It is the first time we have used such an elaborate procedure to ascertain
the accuracy of our results.
All results presented from now on are from
calculations with the direct expansion onto $B$-splines, which is more
efficient both in computer space and time, when the scale of the
simulation increases.  We have compared the results of both methods
for all intensities at 3300 nm (13-photon range) and some intensities
in the 12-photon range (2880 nm).  The convergence of all other
calculations presented was ensured within the $B$-Splines propagation
method only, using well documented techniques~\cite{cormier:1997:ati}:
variation of box-size, $B$-Splines density and order, and variation of
the number of angular momenta. 


\subsection{Behavior of ion yields as a function of intensity}

\begin{figure}
  {\epsfxset \epsfbox{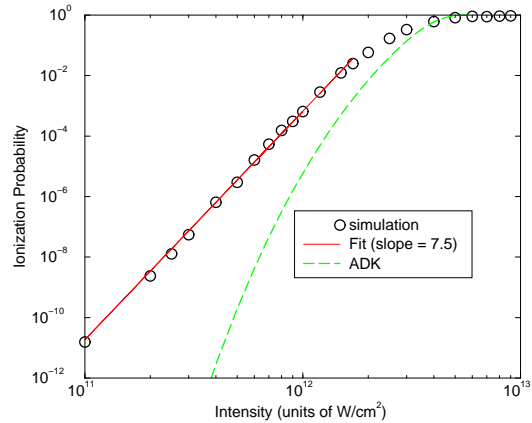}}
  \caption{The ion yield versus intensity, and a fit of the
    low-intensity data to a power law.  A 2880 nm, 96 fs,
    cos $^2$ pulse is used.  The dashed line shows the ADK-theory
    prediction.}
  \label{fig:yield:2880nm:fit}
\end{figure}
We begin the presentation of our results with a study of ion
yields, beginning with
the 12-photon
process at a 2880 nm pulse, of 96 fs temporal width, and
intensities ranging from $10^{11}$ W/cm$^2$ to $10^{13}$ W/cm$^2$.  A
2000 a.u.\ box with 2000 linearly sampled $B$-Splines of order 7, for
each angular momentum up to $l=20$, proves sufficient for obtaining
the ion yield.  In figure~(\ref{fig:yield:2880nm:fit}) we present the
results, together with a power-law fit to the low-intensity part of
the spectrum.  In the same figure we also show the ionization yield
estimate obtained by the ADK-theory: in this and the following
figures where the ADK predictions are displayed, we have integrated
the ADK-rate over a square pulse of maximum intensity equal to the
FWHM of the pulse used.
Saturation sets in at about $3\times 10^{12}$ W/cm$^2$,
substantially higher than the scaled estimate of $10^{11}$ W/cm$^2$,
and in very good agreement with the ADK prediction.
The low-intensity spectrum seems
to follow a multiphoton perturbative behavior, in agreement with the
value of the tunneling parameter $\gamma$ in
table~(\ref{tab:parameter:all}), and thus the power-law; but the least
squares fit yields a slope of 7.5, substantially smaller than the
lowest-order perturbation theory expectation of 12.  The same behavior
appears in the experimental results~\cite{sheehy:1998:ema}, where in
the 12-photon ionization curve, a slope of 7  has been measured.  Note
that the ADK-theory, which is often used in comparisons to
experiments due to its simplicity, markedly fails to predict the
low-intensity yield.

\end{multicols}

\begin{figure}
  {\epsfxsize=16cm \epsfbox{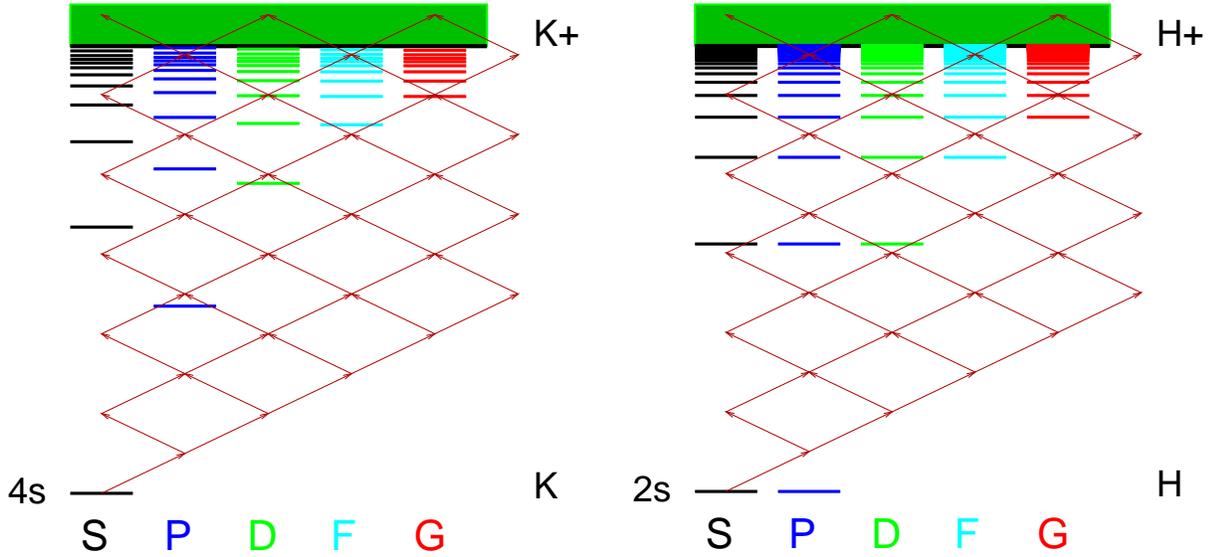}}
  \caption{Potassium under 2880 nm radiation (left), and Hydrogen starting
    from the 2s under 4090 nm radiation (right).  A truncated part of
    the atomic structure is shown, together with the quantum paths
    leading to ionization.}
  \label{fig:compare:K:H2s}
\end{figure}

\begin{multicols}{2}
A scaled system, namely the 12-photon ionization process in Hydrogen
starting from the metastable 2s state and interacting with 4090 nm
light, is used as a test for these results.  These two different systems
are compared in figure~(\ref{fig:compare:K:H2s}), where we show bound
states of Potassium and Hydrogen for the 4 lowest angular momenta,
together with the 12-photon quantum paths leading to ionization, in
energy units scaled to the photon energy.  The 2s state is chosen
instead of the ground state of Hydrogen as the initial state, so that
the ionization threshold energy and the distribution of the bound
states, other than the degeneracies, resemble those of Potassium.

\begin{figure}
  {\epsfxset \epsfbox{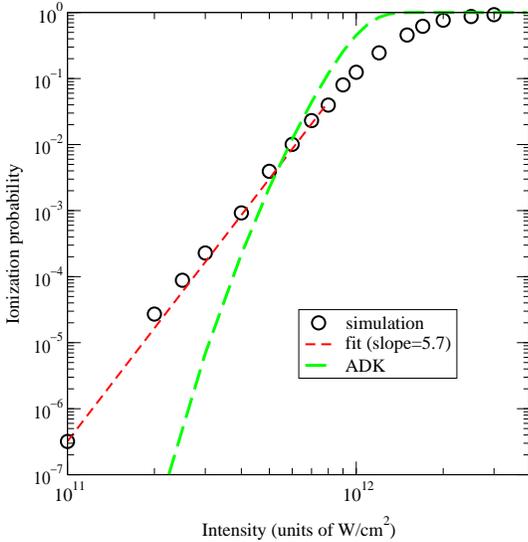}}
  \caption{Ion yield for Hydrogen starting from the 2s in the
    12-photon regime, and a fit of
    the low-intensity part to a power law.   A 4090 nm, 136 fs, cos$^2$
    pulse is used.  The dashed line shows the ADK-theory prediction.}
  \label{fig:H2s:yield:2880-scal}
\end{figure}
\vspace{1cm}
The resulting ion yield for a 136 fs pulse is shown in
figure~(\ref{fig:H2s:yield:2880-scal}).  A 3500 a.u.\ box with 3500
linearly sampled $B$-Splines of order 7 for each of the 21 angular
momenta is used; this is certainly an overkill just for obtaining the
ionization spectrum, but we also analyzed the ATI spectra which need
such large boxes due to the long propagation times involved.  We see
the same kind of structureless saturated spectrum; albeit this time
the yield is higher for the same intensities (notice the scales in the
figures), and the saturation intensity is smaller by a factor of 2, in
accordance with the scaling relations that predict a higher
cross-section for Hydrogen starting from the 2s.\modify{You can
  mention Lars here} 
The ADK theory (shown as the dashed line in the figure) departs at the
lower part of the spectrum, and predicts a smaller saturation intensity.
The low-intensity part again shows a linear
dependence in the log-log plot, and this time the slope is 5.7, even
less than what it is in Potassium.  The slopes in both Potassium and
Hydrogen 2s, roughly equal the order needed to ionize from the first
exited state, the 4p and 3s or 3d respectively, yet, no unambiguous
model conforming to all of our data could be constructed.  Working in
a related context, Pont et al.~\cite{pont:1990:lft} have
constructed a theory describing multiphoton ionization in a strong
field of low frequency $\omega$, obtaining an asymptotic expansion of
the ac quasienergy in powers of $\omega^2$.  Their paper contains
results for the rate of ionization from the 1s state of hydrogen in a
circularly polarized 1064 nm field.  When translated into a log-log
plot of the rate vs intensity, the rate seems to increase roughly like
the eighth power of the intensity, instead of the twelfth power as
should be expected if the rate was perturbative.


\begin{figure}
  {\epsfxset \epsfbox{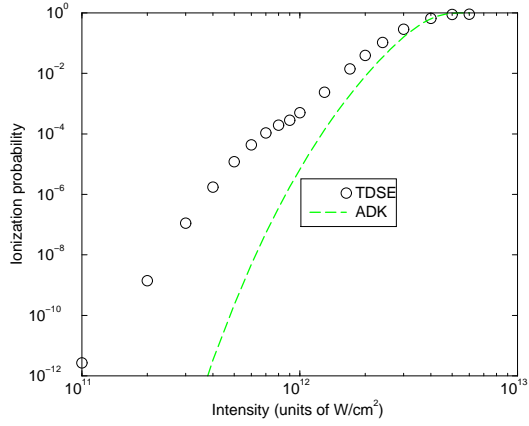}}
  \caption{Ion yield versus intensity.  A 3300 nm, 110 fs,
    cos$^2$ pulse is used.  The dashed line shows the ADK-theory
    prediction.}
  \label{fig:yield:3300nm}
\end{figure}
We move on to 13-photon ionization, where the spectrum shows an
unexpected feature.  In figure~(\ref{fig:yield:3300nm}) we show the
ion yield versus intensity, using 3300 nm, 110 fs pulses, 
calculated within a box of 3000
a.u.\ with 3000 $B$-Splines per angular momentum up to $l=20$; for
comparison we also display the predictions of the ADK theory.  The
saturation intensity is similar to the one in the 12-photon case, and
again a power-law behavior of the signal with intensity holds for the
lowest intensities.  We observe, however, a ``knee'' in
the ion-yield curve, for intensities around $8\times10^{11}$ W/cm$^2$.
This feature resembles a dynamic resonance, rather
unexpected given the very high order of the processes involved.
\end{multicols}

\begin{figure}
\begin{center}
  {\epsfxsize=14cm \epsfbox{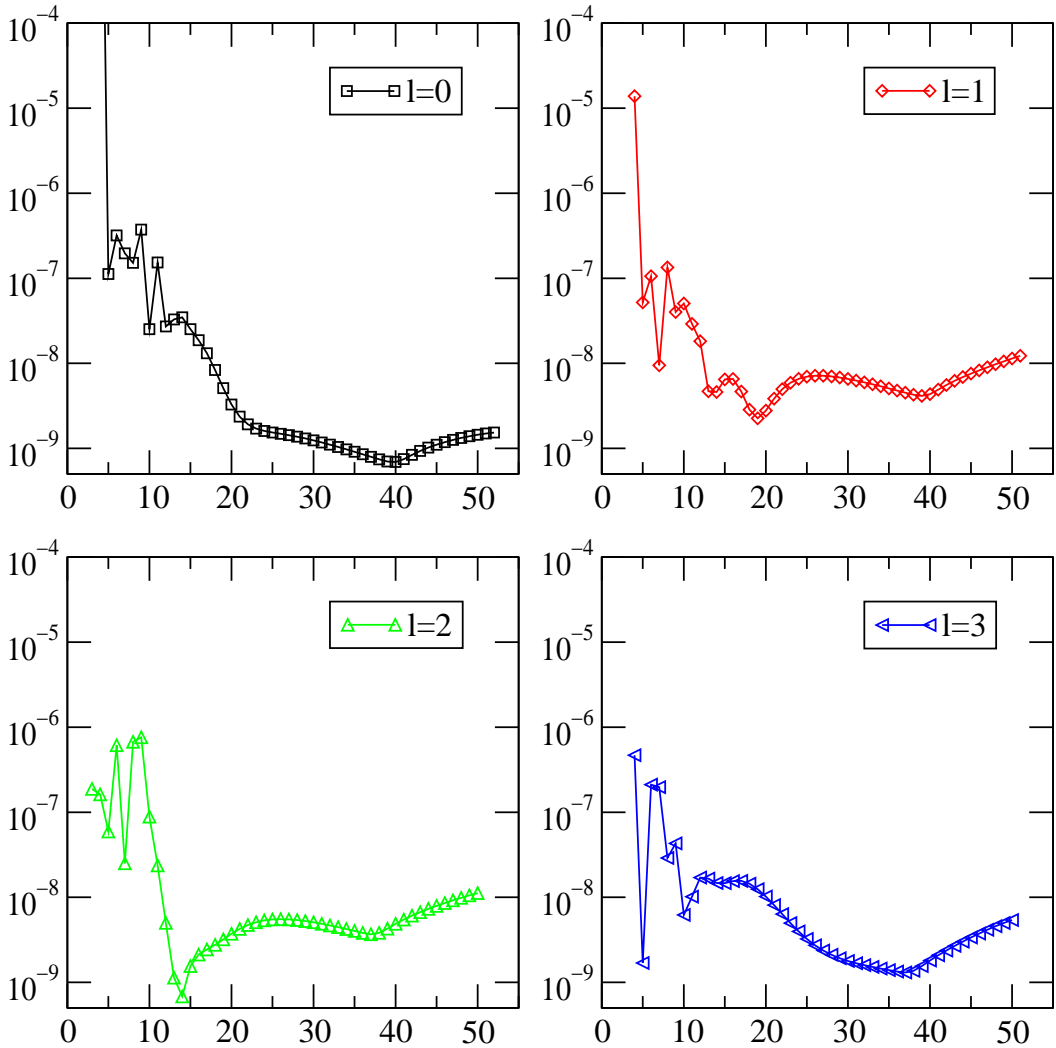}}
\end{center}
  \caption{Bound state distribution at the end of a 3300 nm, 110 fs, cos$^2$
    pulse.  We present the population probability of bound states with
    respect to their main quantum number, for the four lowest angular
    momenta $l$.  The lines have no physical meaning, and are only meant
    as a guide to the eye.}
  \label{fig:bound:dist}
\end{figure}

\begin{multicols}{2}
In figure~(\ref{fig:bound:dist}) we plot the distribution of the final
bound states for an intensity of $7\times10^{11}$ W/cm$^2$.  We plot
the population probability, versus the principal quantum number for a
few of the lowest angular momenta.  Most of the population is still in
the ground state at this intensity; most of the excited
population is concentrated in the $4p$ state, and this is true for all
lower intensities.   Structure appears in the low-angular
momentum, low-excited states, then a smooth, flat part, and a
bump at the highest excitations.  This bump turns out to be
artificial, created by the pseudostates lying between the true bound
states and the discretized continuum states.  This was confirmed by
observing that by making the box smaller, and thus changing the number
of supported bound states, the same effect always appeared in the
region just before the continuum.  The smooth region just before the
bump is expected, given the small energy differences of these states,
compared with the shifts experienced during the pulse.

\begin{figure}
  {\epsfxset \epsfbox{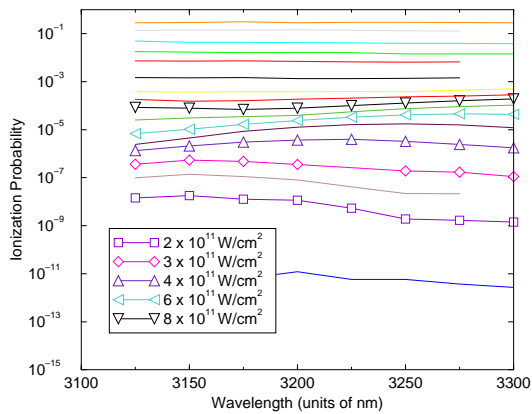}}
  \caption{The ion yield versus wavelength is shown.  A 20 cycles
    cos$^2$ pulse is used.  We use boxes of 2000 a.u., with
    2000 linearly sampled $B$-Splines of order 7, for each of the 21
    lowest angular momenta.}
  \label{fig:yield:scan:13:wl}
\end{figure}
In order to shed more light on the behavior of the ``knee'', we
performed a series of calculations, for different wavelengths,
scanning the entire 13-photon range, from 3125 nm, to 3300 nm, with a
step of 25 nm, and with pulses of 20 optical cycles, whose widths
correspond to 104 fs for the 3125 nm wavelength, and to 110 fs for 3300 nm.
In figure~(\ref{fig:yield:scan:13:wl}), we plot the ion yield as a
function of the wavelength, for a series of intensities ranging from
$10^{11}$ W/cm$^2$, to $3\times10^{12}$ W/cm$^2$.  The curves separate
in a natural way, since for a fixed wavelength a higher intensity
yields more ionization, and we note that for the highest intensities,
above about $10^{12}$ W/cm$^2$, all wavelengths result in practically
the same yield.  In the graph we label only the intensities of
interest 
that are in the range of 2--8$\times 10^{11}$ W/cm$^2$, where
we see a broad resonance shifting to higher wavelengths with
increasing intensity; this suggests a resonance shifting downwards by
increasing the intensity.

\begin{figure}
{\epsfxsize=16cm \epsfbox{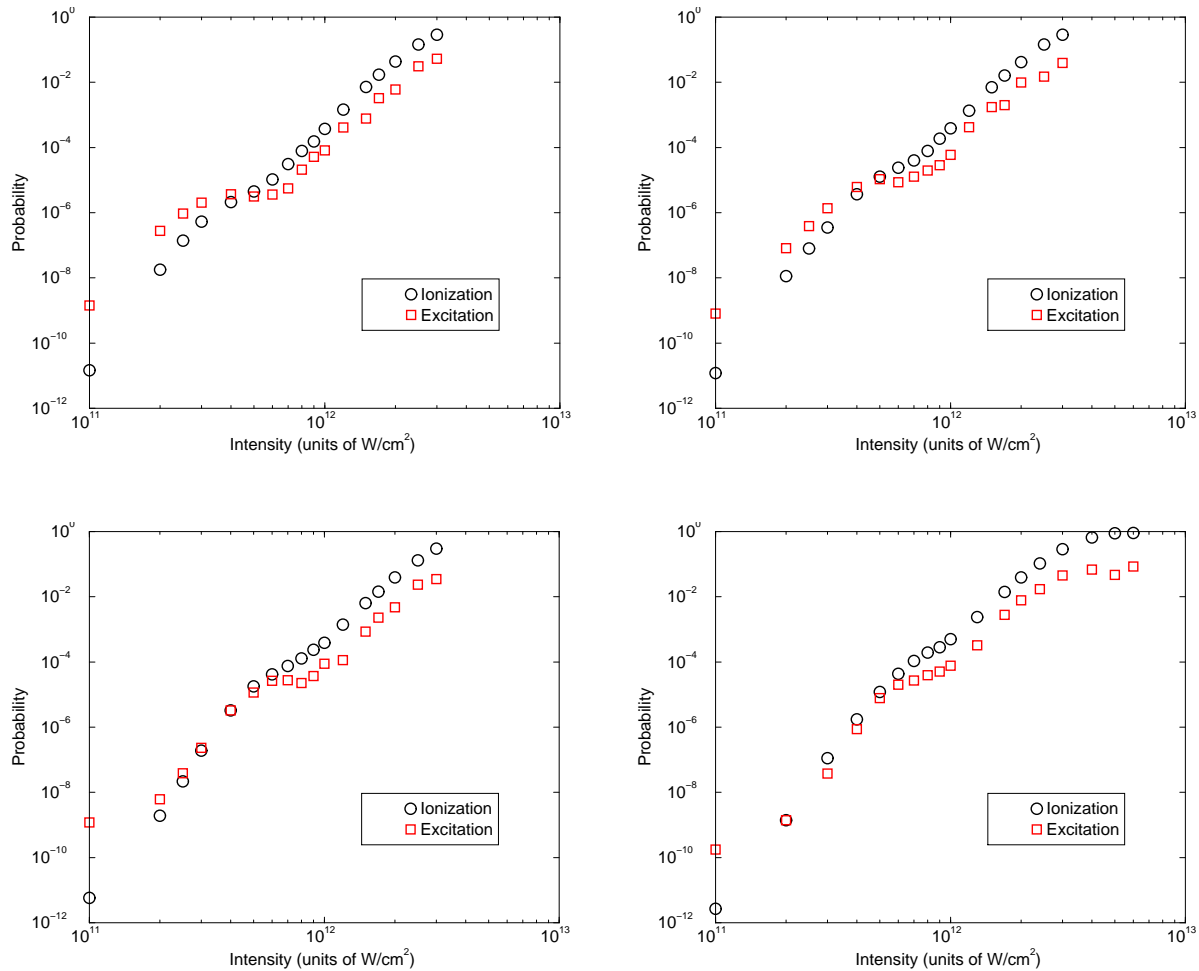}}
  \caption{The ion yield and the total excitation for (left to right,
    and top to bottom) 3150, 3200, 3250, and 3300 nm radiation and a
    20 cycles pulse.  A box of 2000 a.u. is used.}
  \label{fig:ion-exc:3150-3300}
\end{figure}

In figure~(\ref{fig:ion-exc:3150-3300}) we plot (for a few, selected
wavelengths) the ionization curves, together with the total
excitation, which is defined as the population in all bound states
other than the ground state at the end of the pulse.  The excitation
is essentially dominated by the population of the 4p state, for most
of the low- to mid-intensity range.  Note the smooth variation of the
curves with the wavelength; the results for the other wavelengths of
figure~(\ref{fig:yield:scan:13:wl}), essentially interpolate the ones
shown here.  All curves exhibit a ``knee'' which is more pronounced
in the excitation spectrum; after that, ionization mimics excitation
in its behavior, whereas for the lower intensities --- see especially
3150nm and 3200 nm --- their behavior may differ substantially.  One
can argue that the low intensity part is typical in the multiphoton
picture, where the difference in the orders of the processes involved
is expected to show up as a difference in excitation compared to
ionization.  In a Floquet picture, few selected avoided crossings
would describe the bulk of the dynamics.  As the intensity increases,
and higher excitations play a more important role, a regime is
reached, where ionization and excitation are linked, similar to the
transition to the classical chaos regime, which has been discussed in
the microwave context, for example in~\cite{blumel:1987:mih}.
\end{multicols}

\begin{multicols}{2}

For all wavelengths, and all intensities below the ``knee'', the bound
state distribution is qualitatively similar to the one in
figure~(\ref{fig:bound:dist}), with the 4p being by far the most
populated excited state.  On the basis of
figure~(\ref{fig:level-photon:3300nm}), this could happen assuming the
shifts of the 4s and 4p to be such that the two states are brought
closer together when the field is on.  Indeed, by calculating the
lowest order shift in the presence of the field and at the wavelengths
of interest, it turns out that the 4s state shift is negative, and
relatively small, whereas the 4p state, repelled by 5s, shifts down by
more than three times as much.  Thus, for the larger energy
photons in the 13-photon range, a resonant excitation of the
4p state during the pulse occurs; for the lower energy photons of the
same range, the shift pushes the states towards each other easing a
near-resonant transfer.  The phenomenon of a low excitation playing an
essential role in a 13-photon ionization process, is easier to imagine
in an atom like Potassium, than in Hydrogen starting from the ground
state, since the excitation lies at less than half the energy that
is needed to reach ionization.  It should be noted here that
13-photon ionization simulations in Hydrogen, starting from the 2s at
a scaled 13-photon wavelength of 4474 nm, display no corresponding
characteristic, which can be attributed to the difference in the lowest
excitation, as figure~(\ref{fig:compare:K:H2s}) shows.


\begin{figure}
  {\epsfxset \epsfbox{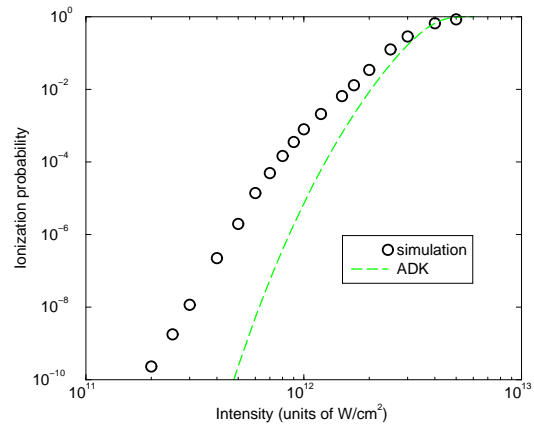}}
  \caption{The ion yield for a 14-photon ionization process.  A 3510
    nm, 117 fs, cos$^2$ pulse is used.  The dashed line shows the ADK
    theory prediction.}
  \label{fig:yield:3510nm}
\end{figure}
We close our discussion of ionization curves, by presenting in
figure~(\ref{fig:yield:3510nm}) the results of 14-photon ionization
calculations, at a 3510 nm wavelength, with a 117 fs pulse.  A 3000
a.u.\  box was used, with 3000 $B$-Splines of order 7 per angular
momentum, up to angular momentum $l=20$.  We also present the ADK
theory predictions: the departure at the lower intensities is not so
dramatic as it was at the shorter wavelengths.


\subsection{Photoelectron energy spectra and ATI}
\label{sec:sub:photoelectron_ATI}

\begin{figure}
  {\epsfxset \epsfbox{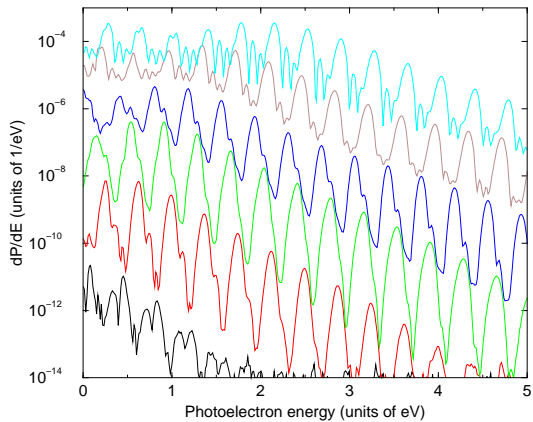}}
  \caption{ATI spectra for (bottom to top) 1, 2, 3, 4, 6, $9 \times
    10^{11}$ W/cm$^2$. 
    A 110 fs, 3300 nm, cos$^2$ pulse is used.}
  \label{fig:ATI:plateau:3300nm:all}
\end{figure}

We move on to the presentation of the ATI spectra, which, as seen in
figure~(\ref{fig:ATI:basis-bsp:compare:7e11}), exhibit a clean plateau
in the low energy range.  This plateau shows up at all wavelengths
we have checked, in Potassium as well as in the Hydrogen simulations
starting from the 2s, 
and it may thus be considered a global feature for mid-infrared
wavelengths.  In figure~(\ref{fig:ATI:plateau:3300nm:all}) a series of
ATI spectra at 3300 nm, 110 fs $\cos^2$ pulses, and for selected 
intensities between $10^{11}$ W/cm$^2$ and $10^{12}$ W/cm$^2$ is shown; the
higher the intensity, the higher the signal shown in the figure.  A
box of 3000 a.u., with 3000 $B$-splines of order 7 for each of the
angular momenta up to $l=20$ is used.  The vertical axis in the figure
corresponds to the ionization probability density in units of 1/eV.
The shift of the 
ATI peaks to lower energies with increasing intensity is linear with
intensity, and, as expected, is well described by the ponderomotive
shift of the ionization threshold.  We note that the extent of the
plateau grows, almost in proportion to the intensity.

%
%

\begin{figure}
  { \epsfxset \epsfbox{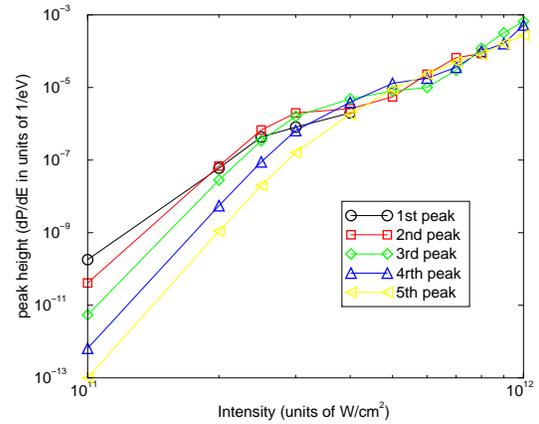}}
  \caption{The height of the first few ATI peaks as a function of
    intensity.  A 3150 nm, 100 fs, cos$^2$ pulse is used.  The lines
    have no physical meaning and are only used as a guide to the eye.
    We notice the convergence of the peaks which indicates the plateau
    formation.}
  \label{fig:ATI:plateau:peaks}
\end{figure}
A clearer way of viewing this phenomenon is presented in
figure~(\ref{fig:ATI:plateau:peaks}).  We plot the height of the first
few ATI peaks versus the intensity of the pulse, this time for a 3150
nm wavelength and a width of 105 fs, corresponding again to a 20-cycle
pulse.  Note the power law (linear dependence on a log-log 
plot) of the first few peak-heights with the intensity.  This
indicates that a perturbative process is taking place.  As the
intensity increases, more peaks enter the plateau range and the
perturbative picture ceases to hold.  This is demonstrated in 
figure~(\ref{fig:ATI:plateau:peaks}) by the
merging of the curves, which implies that for a range of peaks their
heights are basically the same.  As the intensity increases, the
plateau expands and more curves merge.

%
%
\vspace{1cm}
\begin{figure}
  {\epsfxset \epsfbox{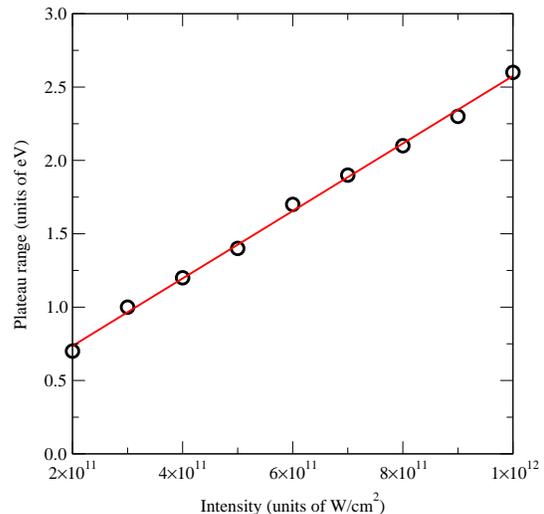}}
  \caption{ATI spectra plateau range as a function of intensity.  A
    3300 nm, 110 fs, cos$^2$ pulse is used.  Hand-read definition of
    plateau range.}
  \label{fig:ATI:plateau:3300nm:fit}
\end{figure}
In figure~(\ref{fig:ATI:plateau:3300nm:fit}), we show the plateau
range plotted versus the intensity for the calculation shown in
figure~(\ref{fig:ATI:plateau:3300nm:all}).  In this figure, the
plateau range is defined as the abscissa of the interception of a
horizontal line joining the first few peaks, and a line over the
decreasing peaks of the spectrum.  The fit was made by hand, to
manually remove the effect of accidental resonances in the first peaks
of the spectrum.  We notice the sharp linear dependence of the plateau
range on the intensity.  On the same figure, we show a fit of the
selected points to a linear function of intensity.  The least squares
fit gives a plateau range of $(2.3 \times 10^{-12} \pm 4 \times
10^{-14})$eV cm$^2$/W$\times I + (0.28 \pm 0.03)$eV.  Measured in
units of 
$U_p$ for the 3300 nm wavelength, the same equation reads: $(2.4 \pm
0.04) U_p + (0.28 \pm 0.03)$ eV.  This shows a plateau scaling in
proportion to $2.4 U_p$.

%
%
\end{multicols}

\vspace{1cm}
\begin{figure}
  {\epsfxsize=17cm \epsfbox{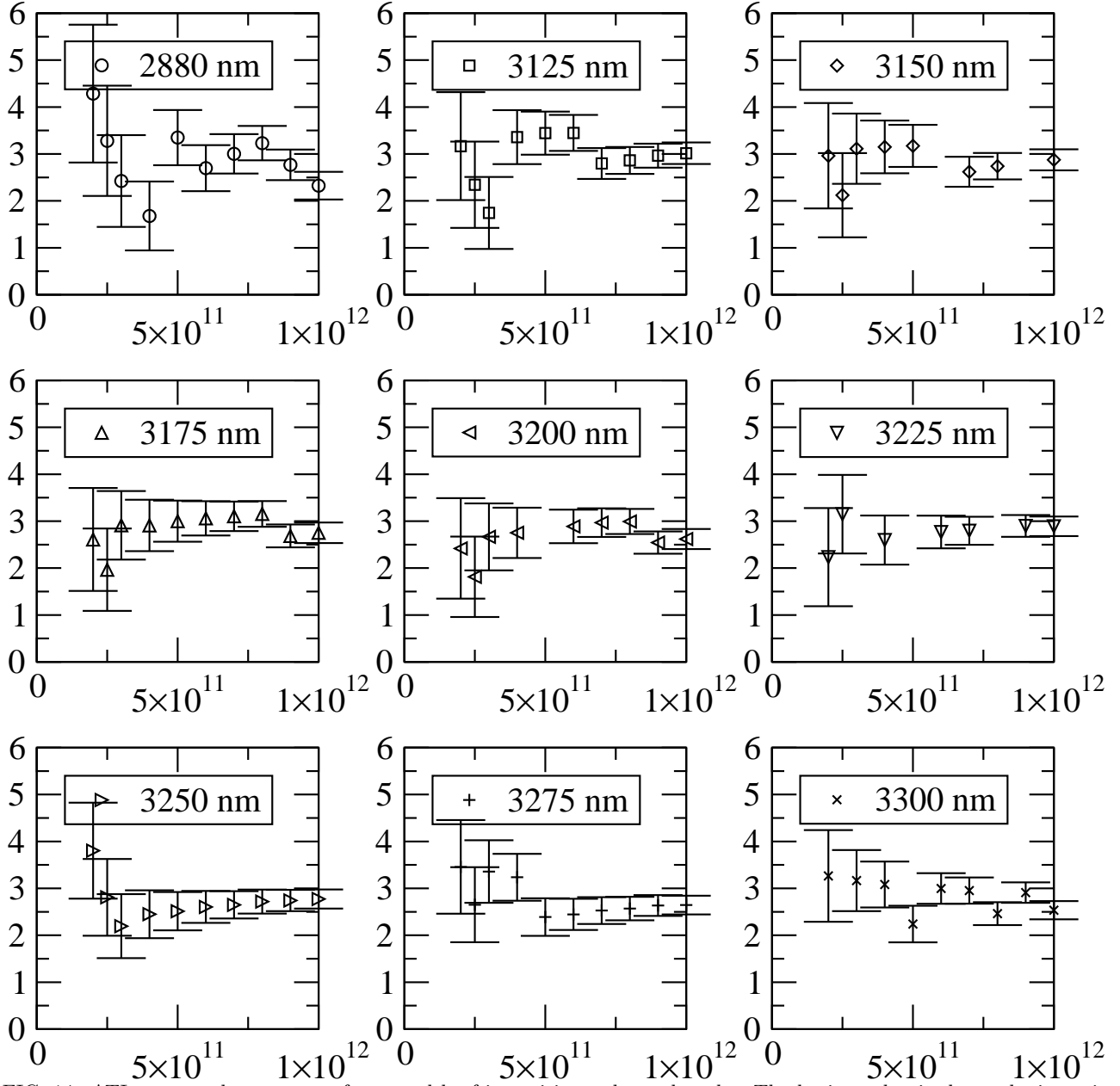}}
  \caption{ATI spectra plateau range for a wealth of intensities and
    wavelengths.  The horizontal axis shows the intensity in units of
    W/cm$^2$; the vertical axis shows the extent of the plateau in
    units of the ponderomotive energy.  The large error-bars present
    in the lowest intensities originate from the rigorous definition
    of the plateau range described in the text.}
  \label{fig:ATI:plateau:all}
\end{figure}

\begin{multicols}{2}
To clarify the relation of the plateau to the ponderomotive energy, we
combine all of the data from the simulations in the 12-, and 13-photon
range.  We simplify the definition of the extent of the plateau to a
rigorous one, whose extraction from the data is automated, namely, we
use the abscissa of the first peak at which the exponential fall of
the peak heights begins.  The results are shown in
figure~(\ref{fig:ATI:plateau:all}), where for each intensity the
plateau range measured in units of $U_p$ is shown for all wavelengths.  The
error-bars reflect the imprecision due to cases in which the plateau
ends between two consecutive peaks.  From the definition of the
plateau range that we use, it follows that the real plateau has the
same probability to be located anywhere within our error-bars.  Notice
the huge uncertainty (larger than $U_p$) at small intensities which
reflects the fact that $U_p$ is smaller than the peak spacing.  At
higher intensities, the ponderomotive shift grows until it is
considerably larger than the peak-spacing, thus shrinking the error-bars
to less than $U_p$.  The data is fitted to a normal
distribution with an average value of 2.8 $U_p$ and a standard
deviation of 0.5 $U_p$.

%
%

This result relates to the classical theory of the
ionization process, as has been first developed in the simpleman's
model~\cite{heuvell:1988:lce,gallagher:1988:ati,corkum:1989:ati}.  The
main arguments have recently been clearly restated in the appendix
of~\cite{lagattuta:1998:qmh}, although in the context of a
rescattering picture~\cite{walker:1996:ers}.  The idea rests upon
a free-electron maximally gaining 3.17 $U_p$ within a
field, returning to its original position.  This happens within the
first cycle, or the first period of the electronic motion, subsequent
cycles lead to 2.4 $U_p$, and then to less until the maxima gradually
converge to 2 $U_p$ over many cycles, and the average kinetic energy
becomes $U_p$.  A relevant study has been made more than 10 years ago
by Gallagher in the microwave regime~\cite{gallagher:1988:ati},
where ionization of Na Rydberg
states was studied, and the spectra were explained using the above
mentioned theory.  In that paper, a plot of the extend of the spectrum
starting from the 40 s state of Na, shows approximate scaling with
3.45 $U_p$.  An energy of $U_p$ should be subtracted when compared to
short-pulse experiments, as the electrons do not keep the
extra ponderomotive energy since they cannot sample the spatial
gradient of the field.  After the observation, in the optical regime,
of the long-range plateau extending from 2 to 8
$U_p$~\cite{paulus:1994:pat,hansch:1997:rhe}, the theory has been
extended to include backscattering from the nucleus, having thus 
provided
answers to pertinent questions~\cite{walker:1996:ers,hu:1997:pat}.  It
should be noted that other than the original Gallagher experiment, all
of these experiments and theories exhibit a steep fall in the region up
to $2U_p$, which then stabilizes or falls with a lower slope in the
region from 2 to 8 $U_p$.  This is in contrast to our findings in the
mid-infrared regime that show a smooth, almost flat region in the
low energy spectra, and an increased downward slope afterwards.  Due
to the numerical demands on convergence, in the present study we do
not analyze the region up to and beyond 8 $U_p$, and cannot therefore
establish the presence or not of a second plateau.

\begin{figure}
  {\epsfxset \epsfbox{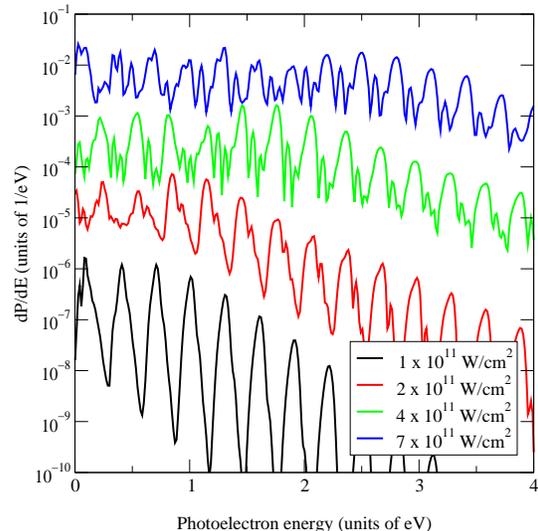}}
  \caption{ATI spectra of Hydrogen starting from the 2s, in a 4090 nm
    field.  The spectra for 1, 2, 4, and $7\times10^{11}$ W/cm$^2$ are
    shown.  A box of 3500 a.u.\  with 3500 $B$-Splines per angular
    momentum is used.}
  \label{fig:H2s:ATI:1-2-4-7e11}
\end{figure}
We have also confirmed that the plateau characteristic has no relation to the
atomic structure involved, by examining the ATI spectrum in the scaled
problem, namely 12-photon ionization of Hydrogen starting from the 2s
state.  The data are from the simulations corresponding to the
wavelength displayed in figure~(\ref{fig:compare:K:H2s}), and the
parameters used are the same as those used to plot
figure~(\ref{fig:H2s:yield:2880-scal}).  Few selected ATI spectra are
shown in figure~(\ref{fig:H2s:ATI:1-2-4-7e11}), all of which display the
plateau, whose range is estimated as $2.6 \pm 0.3 U_p$, close
to what we obtained from the cumulative data in Potassium.



\section{Conclusions}
\label{sec:Conclusions}

In summary, we have conducted a theoretical study of the dynamics of
Potassium interacting with a high intensity, mid-infrared, short,
laser pulse.  
We chose Potassium because it has 
recently been studied experimentally by Sheehy et
al.~\cite{sheehy:1998:ema}.  We have performed time dependent
calculations in terms of 
both a direct expansion of the 
Schr\"odinger equation onto $B$-Splines, as well as 
an expansion onto
field-free eigenstates within a box, and obtained remarkable agreement
between the two methods.  
We have studied a 12-photon ionization process, and have observed that
the low-intensity, power-law 
behavior of the ion yield has an exponent much lower than the
perturbative expectation of 12.  
We have obtained the same behavior from 
the study of a scaled system, namely Hydrogen starting from the 2s in 
the 12-photon range.
In the 13-photon ionization of Potassium, we noted a ``knee''
structure in the ion yield, and linked it to a similar, more
pronounced, behavior in the excitation.  
We interpret this feature
as a broad dynamical resonance with the lowest excited state.  
The ATI
spectra, in all cases that we have studied, display a clean formation
of a plateau in the first few peak-heights.  Although the extension of
a plateau in the ATI spectrum at optical wavelengths from 2 to 8 times the
ponderomotive energy has been discussed in the literature, the
existence of a clear low-energy plateau has (to the best
of our knowledge) not been observed in other studies in the optical or
UV regime, but has been measured in the microwave regime and interpreted
by the simpleman's theory of ionization~\cite{gallagher:1988:ati}.
Through the analysis of the cumulative data from our simulations, we
have determined the extent of the plateau to scale with the
ponderomotive energy $U_p$ as $(2.8 \pm 0.5) U_p$, which is compatible
with the predictions of the classical theory. 
Given the much longer wavelength and therefore longer optical period,
an initially launched wavepacket will have more time to spread before
it backscatters from the core.  This raises the question as to whether
backscattering would play an important role at this wavelength.  Recall
that backscattering for shorter wavelengths has been associated with
changes of the slope of the ATI spectrum up to around 10 $U_p$.  A
related question of course is whether the initially launched wavepacket
might be narrower as a result of which their might not be substantial
additional spreading by the time it reaches the core.  A definitive
evaluation of this aspect would require much more extensive
calculations which may be worth undertaking in the future.

\section{Acknowledgements}

We would like to thank Dr.~L.~F.~DiMauro for making available their
experimental results~\cite{sheehy:1998:ema} prior to publication.
Computer time, space and support from the Rechnenzentrum Garching, and
especially John Cox and Dr.~R.~Volk is gratefully acknowledged.  One
of us (P.~M.)  would like to thank Dr.~L.~A.~A.~Nikolopoulos for useful
discussions.

\section{Appendix}


In the appendix we discuss the model potential used, by comparing it
to the accepted structure of Potassium and to the alternative frozen
core Hartree-Fock model.  The atomic structure of Potassium can be
found in several places.  Since we cannot in a straightforward
manner incorporate relativistic effects in our theory, we use the
weighted energy levels, which are calculated as:
\begin{equation}
  E_{\text{av}}
  =
  \frac{1}{\sum\limits_{J_i}(2 J_i + 1)}
  \sum\limits_{J_i} (2 J_i + 1) E_{J_i}
\end{equation}
The atomic eigenenergies $E_{J_i}$ are taken
from~\cite{smith:1995:asl}.  For easier comparison with the numerical
results, we measure the energies from the ionization threshold which
is at 35009.814 wavenumbers according to Sugar and
Corliss~\cite{sugar:1985:ael}.

The multiplet oscillator strengths are calculated using the
approximate formula~\cite{wiese:1969:atp}:
\begin{equation}
  f_{ik}^{\text{multiplet}}
  =
  \frac{1}{\sum\limits_{J_i}(2 J_i + 1)}
  \sum\limits_{J_i J_k}(2 J_i + 1) f(J_i, J_k)
\end{equation}
The data needed in this formula are taken from~\cite{smith:1995:asl}.
Note that in this database the quantity that is given is $\log((2 J_i
+1) f(J_i, J_k))$.  The multiplet oscillator strengths are shown in
the second column of table~(\ref{tab:Potassium:fosc}).  A few
oscillator strengths are also presented in page 300
of~\cite{sobelman:1992:asr}.  They are in agreement with the ones
shown in the table.
\end{multicols}

\begin{table}
\caption{Potassium energy levels measured from first ionization
  threshold.  E$_{\text{Exp}}$ (cm$^{-1}$) is the weighted
  nonrelativistic value, E$_{\text{HF}}$ is the Frozen Core
  Hartree-Fock result, and E$_{\text{H}}$ is the Hellman Potential
  value.}
\label{tab:Potassium:energies}
\begin{tabular}{dddd}
State             & E$_{\text{Exp}}$ (cm$^{-1}$) &
E$_{\text{HF}}$  (cm$^{-1}$) & E$_{\text{H}}$  (cm$^{-1}$) \\ \hline
$4s$ & -35010 & -32253 & -38947 \\
$5s$ & -13983 & -13376 & -15761 \\
$6s$ & -7560 & -7326 & -8331 \\

$4p$ & -21986 & -20972 & -22647 \\
$5p$ & -10296 & -10000 & -10696 \\
$6p$ & -6005 & -5876 & -6209 \\

$3d$ & -13474 & -12755 & -11952 \\
$4d$ & -7612 & -7213 & -6735 \\
$5d$ & -4824 & -4600 & -4322 \\

$4f$ & -6882 & -6860 & -6852 \\
$5f$ & -4403 & -4390 & -4384 \\
$6f$ & -3057 & -3049 & -3045 \\

$5g$ & -4393 & -4390 & -4389 \\
\end{tabular}
\end{table}

\begin{multicols}{2}
An alternative approach to the model potential, that originally seemed
appealing, is the Frozen Core Hartree-Fock method.  Its main merit is
its {\em ab initio} nature, and the comparatively better description
of atomic structure.  Extensive theoretical discussion exists on the
Frozen Core Hartree-Fock method.  
\modify{[NOTE: A short (one-sentence) historical note is appropriate]}
We used an implementation that is documented in~\cite{chang:1993:bsb}.
The work in that paper was concerned with
the application of
the method to configuration interaction on the Frozen Core Hartree
Fock basis, whereas here we are interested only in the single outer electron
case.  Atomic structure is described quite well, as we see in the
third column of Table~(\ref{tab:Potassium:energies}) where we show
selected bound state energies, and in columns 3 and 4 of
Table~(\ref{tab:Potassium:fosc}), where we present bound-bound
oscillator strengths in the length and velocity gauge starting from s,
p, and d states respectively.  The energies are measured from the
first ionization threshold and are expressed in wavenumbers for direct
comparison with the available data.  A 500 a.u.\ box, with
500 $B$-Splines of order 9 was used for the calculations. 
We notice that the agreement between oscillator strengths calculated
in the length and velocity gauges is quite good for the cases
displayed in the tables.  
The major disadvantage of the method is the inconvenient scaling of the time
needed to calculate the basis, and the actual size of the basis
calculated, which limits us to small cases, up to 1000 a.u..  The
time needed by the frozen core Hartree Fock scales as $N^4$, although
in principle it is limited by a $N^3$ factor.  The size scales with $N^2$
per angular momentum, so that for a $N=3000$ basis of 20
angular momenta, we need approximately 2 GB and 35 CPU days in our
workstation cluster to perform the structure calculations.
Thus it is mainly numerical considerations that dictate the use of
the pseudopotential method.
\end{multicols}

\begin{table}
\caption{Some Potassium multiplet oscillator strengths, starting from few
  lowest $s$, $p$, $d$ states.  f is the exact multiplet oscillator
  strengths, f$_{\text{HF}}$ is the
  Hartree-Fock result calculated in the length gauge, 
  and f$_{\text{H}}$ is the Hellmann potential result.}
\label{tab:Potassium:fosc}
\begin{tabular}{cdddd}
State             & f & f$_{\text{HF}}$(len) & 
f$_{\text{HF}}$(vel) & f$_{\text{H}}$\\ \hline
$4s  \rightarrow 4p $  & 1.01  & 1.07 &        1.02 & 0.96 \\
$4s  \rightarrow 5p $  & 0.009  & 0.01 &  0.008 & 0.026 \\
$4s  \rightarrow 6p $  & 0.0009  & 0.0012 & 0.0008 &0.0055 \\
$5s  \rightarrow 5p $  & 1.49  & 1.52 &         1.49 &  1.42 \\
$5s  \rightarrow 6p $  & 0.031  & 0.026 &     0.024 & 0.075 \\
$6s  \rightarrow 6p $  & 1.92  & 1.96 &         1.94   & 1.82 \\

$4p  \rightarrow 5s $  & 0.18  & 0.18 &        0.17  & 0.19 \\
$4p  \rightarrow 6s $  & 0.019  & 0.018 &      0.017 & 0.009 \\
$4p  \rightarrow 3d $  & 0.89  & 0.93 &       0.97  & 0.90 \\
$4p  \rightarrow 4d $  & 0.0003  & 0.012 &   0.015 & 0.092 \\
$4p  \rightarrow 5d $  & 0.003  & 0.0002 &  0.0005 & 0.028 \\
$5p  \rightarrow 6s $  & 0.31  & 0.32 &        0.31 &0.34 \\
$5p  \rightarrow 4d $  & 1.20  & 1.25 &          1.29 & 1.00 \\
$5p  \rightarrow 5d $  & 0.0077  & 0.032 &    0.036 & 0.14 \\
$5p  \rightarrow 6d $  & 1.48$ \times 10^{-5}$  & 0.0037 &   0.0046 & 0.048 \\

$3d  \rightarrow 5p $  & 0.14  & 0.16 &        0.18  & 0.14 \\
$3d  \rightarrow 6p $  & 0.0066  & 0.0066 &    0.0078 & 6.75$ \times 10^{-9}$ \\
$3d  \rightarrow 4f $  & 0.76  & 0.88 &       0.88 & 1.061 \\
$3d  \rightarrow 5f $  & 0.17  & 0.16 &        0.16 & 0.14 \\
$3d  \rightarrow 6f $  & 0.067  & 0.062 &     0.062  & 0.049 \\
$4d  \rightarrow 6p $  & 0.30  & 0.34 &      0.35 & 0.25 \\
$4d  \rightarrow 5f $  & 0.39  & 0.17 &      0.17 & 0.97 \\
$4d  \rightarrow 6f $  & 0.14  & 0.62 &      0.62 & 0.18 \\
\end{tabular}
\end{table}

\begin{multicols}{2}
The last columns in tables~(\ref{tab:Potassium:energies}),
and~(\ref{tab:Potassium:fosc}) are the results of the Hellmann
pseudopotential, as presented in
equation~(\ref{eq:hellmann:definition}), calculated with a box of 2500
a.u., with 2500 $B$-Splines of order 9.  Since the potential is
$l$-independent, length and velocity gauge oscillator strengths agree,
without the need of introducing a correction to the dipole
operator~\cite{norcross:1973:pc}.  We calculated the standard
deviation of the length and velocity absorption oscillator strengths
starting from a specific state.  For all bound states and up to the
lower part of the continuum spectrum, which is what interests us, this
was less than $10^{-5}$, which reassures us of the completeness of our
description.  The model potential represents the atom less accurately
than the Hartree-Fock does, both energies and oscillator
strengths.  Nevertheless, its excellent numerical properties, originating
from its simplicity, make the large calculations feasible.
Scaling techniques help to map the model atom results onto real
experiments, but our main aim is to study general features pertaining
in the mid-infrared range, thus making the actual atom used of
secondary importance.



\end{multicols}


\end{document}